\documentclass[a4paper,fleqn]{cas-dc}

\usepackage[authoryear,longnamesfirst]{natbib}
\usepackage{lineno,hyperref}
\usepackage{ulem}
\usepackage{algorithm2e,setspace}
\usepackage{subfigure}
\usepackage{multirow}
\usepackage{bbding}
\usepackage{mathrsfs}
\usepackage{amsthm,amsmath,amssymb}
\usepackage{tabularx}
\usepackage{natbib}

\def\tsc#1{\csdef{#1}{\textsc{\lowercase{#1}}\xspace}}
\tsc{WGM}
\tsc{QE}
\tsc{EP}
\tsc{PMS}
\tsc{BEC}
\tsc{DE}

\begin{document}
\let\WriteBookmarks\relax
\def\floatpagepagefraction{1}
\def\textpagefraction{.001}

\shorttitle{}

\shortauthors{Xie et~al.}

\title [mode = title]{DEMONet: Underwater Acoustic Target Recognition based on Multi-Expert Network and Cross-Temporal Variational Autoencoder}                      




%
\author[address1,address2]{Yuan Xie}[
                        style=chinese,
                        orcid=0000-0003-3803-0929]


\ead{xieyuan@hccl.ioa.cn.cn}



\credit{Conceptualization, Methodology, Software, Validation, Formal analysis, Investigation, Data Curation, Writing - Original Draft, Writing - Review \& Editing, Visualization}

\author[address1,address2]{Xiaowei Zhang}[style=chinese]
\ead{zhangxiaowei@hccl.ioa.cn.cn}
\credit{Methodology, Investigation}

\author[address1,address2]{Jiawei Ren}[style=chinese]
\ead{renjiawei@hccl.ioa.cn.cn}
\credit{Data Curation, Supervision}

\author[address1,address2,address3]{Ji Xu}[style=chinese, orcid=0000-0002-3754-228X]
\ead{xuji@hccl.ioa.cn.cn}
\cormark[1]

\credit{Resources, Writing - Review \& Editing, Supervision, Project administration, Funding acquisition}

\affiliation[address1]{organization={Key Laboratory of Speech Acoustics and Content Understanding, Institute of Acoustics, Chinese Academy of Sciences},
    addressline={No.21, Beisihuan West Road, Haidian District},
    postcode={100190},
    city={Beijing},
    country={China}}
    
\affiliation[address2]{organization={University of Chinese Academy of Sciences},
    addressline={No.80, Zhongguancun East Road, Haidian District}, 
    postcode={100190},
    city={Beijing},
    country={China}}

\affiliation[address3]{organization={State Key Laboratory of Acoustics, Institute of Acoustics, Chinese Academy of Sciences},
    addressline={No.21, Beisihuan West Road, Haidian District},
    postcode={100190},
    city={Beijing},
    country={China}}

\cortext[cor1]{Corresponding author}



\begin{abstract}
Building a robust underwater acoustic recognition system in real-world scenarios is challenging due to the complex underwater environment and the dynamic motion states of targets. A promising optimization approach is to leverage the intrinsic physical characteristics of targets, which remain invariable regardless of environmental conditions, to provide robust insights. However, our study reveals that while physical characteristics exhibit robust properties, they may lack class-specific discriminative patterns. Consequently, directly incorporating physical characteristics into model training can potentially introduce unintended inductive biases, leading to performance degradation. To utilize the benefits of physical characteristics while mitigating possible detrimental effects, we propose DEMONet in this study, which utilizes the detection of envelope modulation on noise (DEMON) to provide robust insights into the shaft frequency or blade counts of targets. DEMONet is a multi-expert network that allocates various underwater signals to their best-matched expert layer based on DEMON spectra for fine-grained signal processing. Thereinto, DEMON spectra are solely responsible for providing implicit physical characteristics without establishing a mapping relationship with the target category. Furthermore, to mitigate noise and spurious modulation spectra in DEMON features, we introduce a cross-temporal alignment strategy and employ a variational autoencoder (VAE) to reconstruct noise-resistant DEMON spectra to replace the raw DEMON features. The effectiveness of the proposed DEMONet with cross-temporal VAE was primarily evaluated on the DeepShip dataset and our proprietary datasets. Experimental results demonstrated that our approach could achieve state-of-the-art performance on both datasets.
\end{abstract}




\maketitle

\section{Introduction}
Underwater acoustic target recognition is a crucial component
of marine acoustics~\cite{rajagopal1990target, tian2023joint, jin2023offshore}, aimed at recognizing underwater targets such as ships by analyzing their radiating sound. This technology has extensive applications in underwater surveillance and navigation systems~\cite{sutin2010stevens, jia2022deep}, security defense systems, and in the development and protection of marine resources~\cite{etter2012advanced, fillinger2010towards}.

As a vital task in underwater acoustics, underwater acoustic target recognition is usually accompanied by many challenges in practical scenarios~\cite{xie2022underwater}, such as complex underwater environments, unpredictable transmission channels, and dynamic motion states of targets~\cite{erbe2019effects, xu2023underwater}. In the past, commonly used underwater acoustic features focused on well-defined physical characteristics~\cite{rajagopal1990target, vaccaro1998past}. For instance, peaks and harmonics in DEMON (detection of envelope modulation on noise) spectra, which provide information about propeller shaft frequency and blade counts, have been utilized for underwater target recognition~\cite{wang2021design,liu2020design,hashmi2023novel}, etc. These characteristics are derived from the inherent properties of targets and are less influenced by complex marine environments~\cite{gao2021automatic}, making them interpretable and robust features. However, these features tend to emphasize local physical characteristics, making it challenging to establish a mapping relationship with diverse target categories. In other words, they may not offer sufficient class-specific discriminative patterns to support comprehensive modeling. For example, while DEMON spectra can differentiate between targets with three-blade and five-blade propeller configurations, they may struggle to distinguish between targets that have the same propeller blade counts but belong to different categories, such as passenger ships and cargo ships. Related studies have demonstrated that physical characteristics exhibit limitations in recognition tasks involving diverse categories~\cite{xie2022adaptive} and diverse feature space~\cite{irfan2021deepship}. Additionally, considering the limited capability to handle non-stationary signals, physical characteristics have gradually been overlooked in recognition tasks within this data-driven era.

With the widespread adoption of deep neural networks, time-frequency features, such as spectrograms, have become the prevailing choice for recognition tasks~\cite{yang2019parameterised, li2011extraction}. These features can provide comprehensive information that includes frequency components and time-varying characteristics, thus supporting complex modeling. Previous studies~\cite{xie2022underwater, xie2022adaptive} have reported that recognition systems based on time-frequency features can achieve leading performance on public underwater acoustic datasets~\cite{irfan2021deepship,santos2016shipsear}. However, time-frequency features are susceptible to interferences caused by environmental and motion variations~\cite{xie2022underwater,xu2023underwater}, which may cause recognition models to overfit on target-irrelevant information, such as background noise, channel noise, spurious line spectra, and Doppler shift~\cite{liu2020design}, etc. This vulnerability undermines the generalization and robustness of spectrogram-based recognition systems, thereby limiting their performance in practical application scenarios.

In general, physical characteristics exhibit greater resistance to interferences but lack sufficient class-specific discriminative patterns, whereas spectrograms contain ample class-specific information but are less robust. A question naturally arises: ``Can we integrate physical characteristics into mainstream spectrogram-based recognition systems to compensate for the lack of robustness?'' In our preliminary experiments (see Section 5.1 and Table~\ref{tab_demon}), we attempted to integrate physical characteristics using strategies such as feature fusion~\cite{xie2023guiding} or model ensemble~\cite{yang2016underwater}. However, the experimental results revealed that directly integrating physical characteristics could not yield satisfactory results and, in some cases, even had negative effects. According to the analyses, we attribute the detrimental effect to the lack of class-specific discriminative patterns. Imposing a mapping between physical characteristics and target types can introduce inappropriate inductive biases that hurt the model's performance.

In this study, we propose DEMONet, which leverages DEMON spectra to provide robust insights into shaft frequency or blade counts of targets. To benefit from physical characteristics while avoiding potential detrimental effects, DEMONet incorporates multiple separate expert layers and a routing layer, wherein the routing layer assigns inputs to their best-matched expert layer based on the physical characteristics provided by DEMON spectra. This approach treats physical characteristics as the basis for routing, thus enabling each expert layer to learn from data with common physical characteristics to acquire specialized knowledge. In essence, our design resembles ``clustering before classification''~\cite{mathivanan2018improving}. It facilitates the model's ability to process signals with diverse physical characteristics in a differentiated and fine-grained manner, and prevents the DEMON spectra from directly mapping to target categories. In addition, we leverage the time-invariant nature of physical characteristics to design the cross-temporal variational autoencoder (VAE), which takes the reconstruction of DEMON spectra belonging to the same signal across different periods as the training objective. The reconstructed DEMON spectra are less susceptible to noise and spurious modulation spectra, which are better suited to serve as the input DEMON feature. Additionally, we incorporate the load balancing loss~\cite{fedus2022switch} to address the issue of expert layer underfitting caused by potential load imbalance. To validate the effectiveness of our proposed methods, we conducted extensive comparative experiments and ablation experiments on multiple underwater acoustics datasets. The experimental results demonstrated that DEMONet with cross-temporal VAE can outperform current advanced models on the DeepShip~\cite{irfan2021deepship} and our proprietary dataset~\cite{ren2019feature}. Besides, we investigated the performance of DEMONet on the data-scarce ShipsEar dataset, further clarifying its practical value and application scenarios. The main contributions of this study are summarized as follows:


\begin{itemize}
\item Identification of limitations of physical characteristics and mainstream time-frequency features in underwater acoustic recognition.

\item Proposal of DEMONet, a multi-expert structure that integrates physical characteristics into recognition systems to enhance robustness without the risk of performance degradation.

\item Design of cross-temporal variational autoencoder to mitigate noise and interference in DEMON spectra, along with the utilization of load balancing loss to address the load imbalance issue.
\end{itemize}

\section{Related Works}
Driven by practical applications, underwater acoustic target recognition has been an important research topic in the field of underwater acoustics for numerous years. In early studies, manually designed acoustic features containing intrinsic signal characteristics were usually employed along with classic machine learning techniques. For instance, Li et al.~\cite{xueyao2000application} extracted zero-crossing rate, LOFAR (Low Frequency Analysis Recording) spectrum, and wavelet spectrum to recognize underwater targets; Lu et al.~\cite{lu2020fundamental} employed the DEMON spectrum to analyze the fundamental frequency of signals and realize recognition; Das et al.~\cite{das2013marine} utilized cepstral features with cepstral liftering and Gaussian mixture models (GMMs) for marine vessel classification; Wang and Zeng~\cite{wang2014robust} employed bark-wavelet analysis in combination with the Hilbert-Huang transform to analyze signals, and employed support vector machines (SVMs) as the classifier, etc. However, recent studies~\cite{xie2022underwater, xie2023guiding} have demonstrated that although manually designed acoustic features possess strong physical significance, they often focus on local physical characteristics, lacking a comprehensive perception of signals and sufficient discriminative patterns. Consequently, the recognition system's ability to perform satisfactorily may be hindered when confronted with large-scale datasets encompassing diverse feature spaces~\cite{irfan2021deepship}.

With the development of deep learning techniques~\cite{lecun2015deep} and the accumulation of open-source underwater acoustic databases~\cite{irfan2021deepship,santos2016shipsear}, recognition algorithms based on deep neural networks have gained significant attention. Compared to traditional methods, deep learning methods prioritize acoustic features that encapsulate time-frequency information (e.g., spectrograms), enabling comprehensive modeling of non-stationary underwater signals~\cite{xu2023underwater,xie2023guiding,liu2021underwater}. In the literature, Zhang et al.~\cite{zhang2021integrated} utilized the short-time Fourier transform (STFT) amplitude spectrogram, STFT phase spectrogram, and bispectrum features as inputs for convolutional neural networks; Liu et al.~\cite{liu2021underwater} employed convolutional recurrent neural networks with 3-D Mel-spectrograms and data augmentation for underwater target recognition; Dawid Połap et al.~\cite{polap2023bilinear} proposed bilinear pooling with poisoning detection module and adopted CNN to realize automatic acoustic analysis; Xie et al.~\cite{xie2022adaptive} utilized learnable fine-grained wavelet spectrograms with the deep residual network (ResNet)~\cite{he2016deep} to recognize ship-radiated noise adaptively, etc. In recent years, spectrogram-based recognition models have become the prevailing approach in the field of underwater acoustic recognition due to their superior performance on existing datasets. However, these models possess notable limitations, particularly regarding their robustness in practical application scenarios. They are susceptible to interferences arising from environmental and motion variations. Additionally, their performance may degrade when faced with unseen signals or signals that differ significantly from the training data distribution. Therefore, there is still a pressing need to develop more accurate and robust underwater acoustic recognition systems.

\section{Methodology}

\begin{figure*}[htbp]
    \centering
    \includegraphics[width=0.9\linewidth]{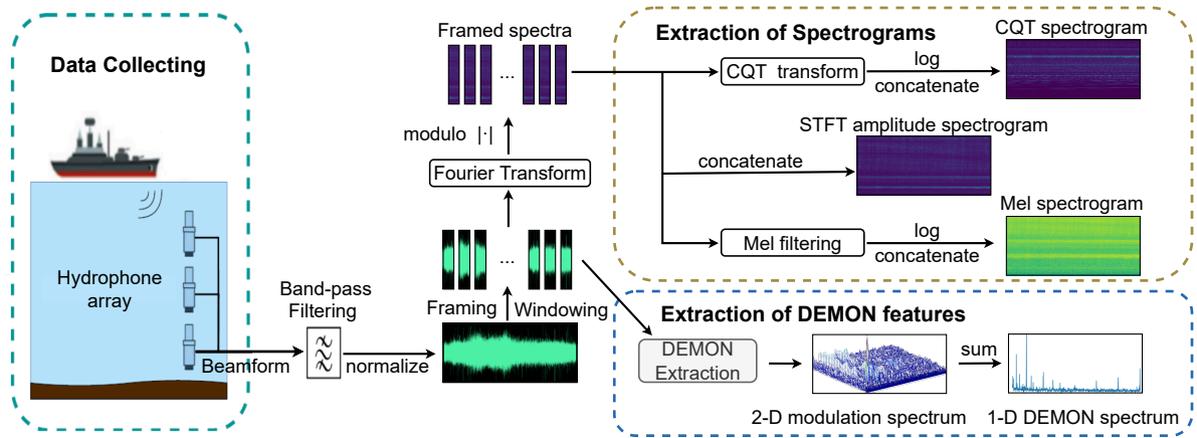}
    \caption{The general process of the data acquisition, preprocessing, and feature extraction.}
    \label{fig1}
    \vspace{-2px}
\end{figure*}

This section is organized as follows: Section 3.1 introduces the preprocessing and feature extraction steps for three candidate spectrograms and the DEMON spectra. Section 3.2 provides a detailed description of the overall framework and the training process of DEMONet with cross-temporal VAE. Section 3.3 presents the detailed architectures of the individual modules of our proposed model.

\subsection{Preprocessing and Feature Extraction}

The preprocessing process is briefly illustrated in Figure~\ref{fig1} (left). Initially, the noise radiated by targets is received by the hydrophone array, after which beamforming algorithms~\cite{castellini2008acoustic} are employed to synthesize single-channel audio signals~\cite{xie2022adaptive}. Then, band-pass filters are applied to single-channel signals to eliminate low-frequency interference and redundant high-frequency components. The specific passbands of used datasets are determined through preliminary experiments (refer to Appendix A), and the parameter setups can be found in Table~\ref{tab1}.

Following the preprocessing stage, feature extraction is performed on the filtered signals. In this process, time-frequency spectrograms and DEMON spectra are jointly extracted to serve as the input for DEMONet. This study adopts three candidate spectrogram features: the STFT amplitude spectrogram, Mel spectrogram, and CQT spectrogram. The detailed extraction process is illustrated in Figure~\ref{fig1} (right). For filtered signals, we begin by applying mean-variance normalization to standardize the waveform amplitude ranges, ensuring they fall within an interval with a mean of 0 and a variance of 1. The normalized signals are then divided into frames with a length of 50 ms and a frame shift of 25 ms. Then, the Hanning window is applied to each short-time frame to mitigate spectrum leakage. Afterward, the short-time Fourier transform (STFT) is performed on the windowed frames to obtain the complex spectra. The modulus of the complex spectra (i.e. amplitude spectra) is then extracted and concatenated across all frames to obtain the STFT amplitude spectrogram.

Next, we apply Mel filter banks to the amplitude spectra for Mel filtering. The Mel filter bank, as illustrated in Equation (1), consists of a set of bandpass filters spaced according to the non-linear Mel scale.

\begin{equation}
\begin{aligned}
    Mel(f)
    &=2595\times log(1+\frac{f}{700}).
\end{aligned}
\end{equation}

The filtered spectra then undergo a logarithmic transformation and are concatenated across all frames to obtain the Mel spectrogram. The Mel spectrogram provides higher frequency resolution in the low-frequency range, which facilitates the analysis of low-frequency line spectrum components. Subsequently, we apply the constant-Q transform (CQT) to obtain the CQT spectrogram. In this process, each frame's amplitude spectrum is convolved with the CQT kernel, which consists of a bank of logarithmically spaced bandpass filters. Among them,  the center frequency component of the $k$-th filter, denoted as $f_k$, is determined by Equation (2):

\begin{linenomath}
\begin{equation}
f_k = 2^{\frac{k}{b}} f_{min},  \qquad k=0,1,\ldots\lceil b\cdot log_2(\frac{f_{max}}{f_{min}}) \rceil -1
\end{equation}
\end{linenomath}

where the octave resolution is represented by $b$, and the upper and lower frequencies to be processed are indicated by $f_{max}$ and $f_{min}$, respectively. The ratio of the filter bandwidth $BW$ to the center frequency is a constant $Q=\frac{f_k}{BW}=\frac{1}{2^{1/b}-1}$. After the filtering, spectra undergo a logarithmic transformation and are concatenated across all frames to obtain the CQT spectrogram. The CQT spectrogram offers higher frequency resolution in the low-frequency range and improved temporal resolution in the high-frequency range. It not only facilitates the analysis of low-frequency components but also provides information on periodic modulation in the high-frequency portion, such as propeller rhythm.

In addition, this study also extracts the DEMON spectra. Firstly, the frequency band of the signal is divided into several sub-bands at intervals of 250Hz (e.g., 10hz-260hz, 260hz-510hz...). The signal is then band-pass filtered according to these sub-bands, yielding signals with different frequency bands. For each filtered signal, assuming that the carrier wave is a single-frequency signal, the modulated signal can be expressed as:

\begin{equation}
    x(t) = A(1+m sin\Omega t)\cdot cos\omega t.
\end{equation}

In Equation (3), $A$ represents the signal amplitude, $m$ represents the modulation depth, $\omega$ represents the carrier frequency, and $\Omega$ represents the modulation frequency. This study employs absolute value low-pass demodulation~\cite{chen2017passive} as the demodulation strategy. The absolute value of the modulated signal $|x(t)|$ can be decomposed as follows:

\begin{equation}
\begin{aligned}
    |x(t)|
    &=A(1+m sin\Omega t)\cdot |cos\omega t| \\
    &=A(1+m sin\Omega t) \frac{4}{\pi}(\frac{1}{2}+\frac{1}{1\cdot3}cos2\omega t-\frac{1}{3\cdot5}cos6\omega t+...) \\
    &=\frac{2A}{\pi} + \frac{2A}{\pi}msin\Omega t+\frac{4A}{3\pi}cos2\omega t +\frac{2mA}{3\pi}sin(2\omega+\Omega)t\\
    &\quad -\frac{2mA}{3\pi}sin(2\omega-\Omega)t+... , \\
\end{aligned}
\end{equation}

which indicates that $|x(t)|$ contains a direct current component ($\frac{2A}{\pi}$), modulation frequency harmonic components related to the modulation degree ($ \frac{2A}{\pi}msin\Omega t$), and other high-order harmonic components. Since $\omega>>\Omega$, the modulation frequency component can be obtained through low-pass filtering. In this study, a second-order zero-phase filter with a Hamming window of window length 1024 is applied for low-pass filtering. Finally, we perform the Fourier transform on the square of the filtered signal, followed by a modulo operation, to obtain the modulation spectrum. The modulation spectra corresponding to each sub-band are then concatenated to form the 2D-DEMON spectrum. To represent the modulation components of the signal more intuitively, the 2D-DEMON spectrum can be summed along the modulation frequency dimension to obtain the 1D-DEMON spectrum. The peaks in the 1D-DEMON spectrum reflect the shaft frequency and blade frequency of the propeller (fundamental frequency and its harmonics), which contain robust physical characteristics of the target.

\begin{figure*}[ht]
    \centering
    \includegraphics[width=1\linewidth]{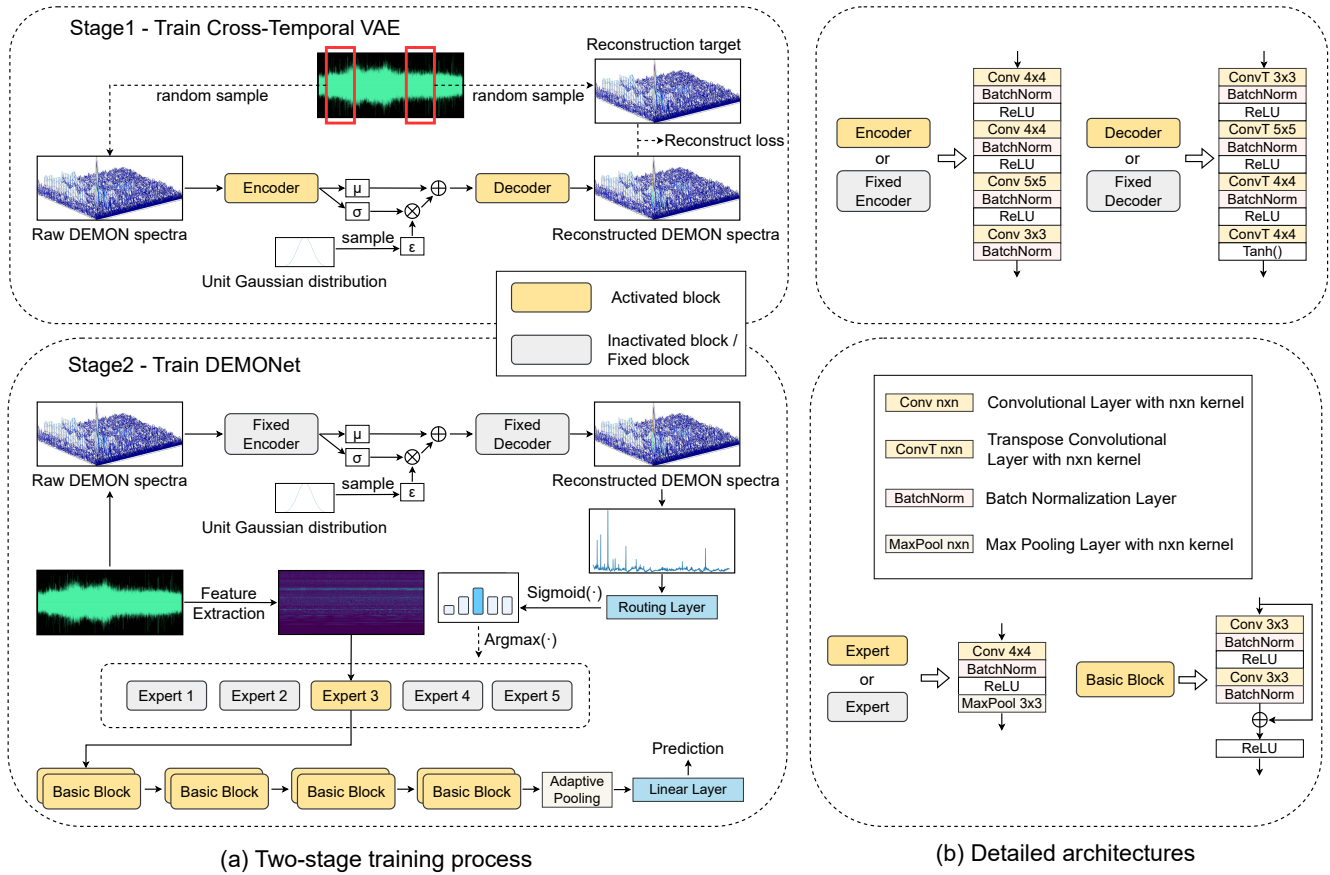}
    \caption{The framework and training process of  DEMONet with cross-temporal VAE, along with the detailed architectures of individual modules of the model. In Figure (a), the orange box signifies the active block, while the gray box indicates that the block is not activated, or the block only performs feedforward without gradient calculation and parameter update.}
    \label{fig2}
    \vspace{-2px}
\end{figure*}

\subsection{Framework and Training Process}

The DEMONet framework is illustrated in Figure~\ref{fig2}. It consists of four main components: the cross-temporal VAE, a routing layer, multiple expert layers, and the backbone network. In the following paragraphs, these components are introduced sequentially in the training process.

The first stage of the training process focuses on training the cross-temporal VAE (see Figure~\ref{fig2} (a) -- Stage 1).  The structure and training process of the cross-temporal VAE closely resemble those of a vanilla VAE. Both models utilize an encoder to convert inputs to the estimated mean ($\mu$) and variance ($\sigma^2$) of latent variables that are assumed to follow the Gaussian distribution, and employ a decoder to reconstruct the target based on the Gaussian latent variables. Drawing inspiration from the time-invariant nature of physical characteristics, the training objective of cross-temporal VAE is redesigned to reconstruct the DEMON spectra belonging to the same signal recording but across different periods. This training objective can encourage the model to focus on time-invariant physical characteristics in the DEMON spectra, such as propeller shaft frequency and blade counts. This helps reduce the impact of noise or spurious modulation spectral lines on the reconstructed DEMON spectra. To train the cross-temporal VAE, we start by randomly sampling two 30-second signal segments, denoted as $wav_A$ and $wav_B$, from a single signal recording $wav$. The DEMON spectrum $d_A\sim \mathbb{R}^{B\times1\times28\times1172}$ ($B$ denotes the batch size, 1 signifies the number of channels, and $28\times1172$ represents the dimension of the 2-D DEMON spectra) corresponding to $wav_A$ serves as the input for the cross-time VAE, while the DEMON spectrum $d_B\sim \mathbb{R}^{B\times1\times28\times 1172}$ corresponding to $wav_B$ is used as the reconstruction target. The encoder takes $d_A$ as input and generates the matrix $v\sim \mathbb{R}^{B\times256\times1\times 287}$, which can be chunked along the channel dimension into two variables: $\mu\sim \mathbb{R}^{B\times128\times1\times287}$ and $log(\sigma^2)\sim \mathbb{R}^{B\times128\times1\times287}$. The use of $log(\sigma^2)$ instead of $\sigma^2$ ensures that the neural network is not constrained to output non-negative values. To obtain the Gaussian latent variables $z$, the model randomly samples $\epsilon$ from the unit Gaussian distribution $N(0,1)$ and computes $z = \epsilon \cdot \sigma + \mu = \epsilon \cdot e^{\frac{1}{2}log(\sigma^2)} + \mu$. The decoder then converts $z\sim \mathbb{R}^{B\times128\times1\times287}$ into the reconstructed DEMON spectrum $\hat{d_A}\sim \mathbb{R}^{B\times1\times28\times 1172}$. The loss function for the cross-temporal VAE, denoted as $\mathcal{L}_\mathrm{VAE}$, is defined as the sum of two terms:

\begin{equation}
\begin{aligned}
    \mathcal{L}_\mathrm{VAE} 
    &=\mathcal{L}_\mathrm{reconstruct} +  \mathcal{L}_\mathrm{KL} \\
    &=MSE(d_B, \hat{d_A}) + KL(N(\mu,\sigma^{2}),N(0,I)). \\
\end{aligned}
\end{equation}

The reconstruction loss $\mathcal{L}_\mathrm{reconstruct}$ is calculated using the mean squared error (MSE) between $\hat{d_A}$ and $d_B$, aiming to minimize the difference between the reconstructed DEMON spectra and the target spectra belonging to another period. The KL divergence loss $\mathcal{L}_\mathrm{KL}$ is used to constrain the Gaussian latent variables $z$ to conform to the standard normal distribution. Eventually, the parameters of the encoder and decoder are updated based on the gradient that minimizes $\mathcal{L}_\mathrm{VAE}$.

The second stage of the training process focuses on training the DEMONet (see Figure~\ref{fig2}(a) Stage 2), which begins after completing the training of cross-temporal VAE. Let $\mathcal{X}=\{x_i\} (i=1,2...B)$ represent the input signals, with $B$ as the batch size. Similarly, let $\mathcal{Y}={y_i}$ denote category labels, $\mathcal{S}={s_i}$ represent input spectrograms, and $\mathcal{D}={d_i}$ represent DEMON spectra. Initially, the DEMON spectra $\mathcal{D}\sim \mathbb{R}^{B\times1\times28\times1172}$ are fed into the pre-trained cross-temporal VAE (the parameters of VAE remain fixed in Stage 2) for forward propagation to obtain the reconstructed DEMON spectra $\hat{\mathcal{D}}=\{\hat{d}_i\}\sim \mathbb{R}^{B\times1\times28\times1172}$. Subsequently, all reconstructed DEMON spectra are summed along the modulation frequency dimension to obtain the 1-D modulation spectra $\hat{\mathcal{D}}_{1d}\sim \mathbb{R}^{B\times 1172}$, which serves as the input for the routing layer. The routing layer $R(\cdot)$ is positioned after the cross-temporal VAE and before the expert layers. It is responsible for assigning input signals to their best-matched expert layer based on the physical characteristics provided by the DEMON features. Specifically, the routing layer takes $\hat{\mathcal{D}}_{1D}$ as input and outputs the routing probability matrix $P=[p_1,p_2...p_B]=\mathrm{Sigmoid}(R(\hat{\mathcal{D}}_{1D}))\sim \mathbb{R}^{B\times N}$, where $N$ represents the number of expert layers. The value of routing probability $p_i\sim \mathbb{R}^{N}$ indicates the probability of the $i$-th input being assigned to each expert layer. The expert layers $\mathrm{Expert}_1(\cdot), \mathrm{Expert}_2(\cdot) ... \mathrm{Expert}_N(\cdot)$ are positioned after the routing layer. These layers have identical structures but different parameters. Among them, only the best-matched expert layer (i.e., with the highest routing probability value) is activated for each input, while the rest of the experts remain inactive. This sparse activation mechanism allows each expert to specialize in learning from data with characteristics in common. Taking Figure~\ref{fig2} (a) as an example, if the routing probability $p_i$ is given as (0.1, 0.2, 0.3, 0.2, 0.2), it signifies that the input feature $s_i$ has a 10\%, 20\%, 30\%, 20\%, 20\% probability of being assigned to $\mathrm{Expert}_1(\cdot), \mathrm{Expert}_2(\cdot), \mathrm{Expert}_3(\cdot), \mathrm{Expert}_4(\cdot), \mathrm{Expert}_5(\cdot)$, respectively. Thus, the input will be assigned to $\mathrm{Expert}_3(\cdot)$ with the highest probability of 0.3. After that, the activated expert layers convert input features into intermediate representations and send them to the subsequent backbone network $F(\cdot)$. The backbone network is responsible for converting the intermediate representations into the prediction probabilities $Z=[z_1,z_2...z_n]\sim \mathbb{R}^{B\times C}$, where $C$ represents the number of categories to be recognized:

\begin{equation}
\begin{aligned}
    z_i &= F(\mathrm{Expert}_{I}(s_i)). \quad I=\underset{j}{{\arg\max} \, p_i[j]}
\end{aligned}
\end{equation}

Finally, the prediction result can be obtained by performing Argmax on $z_i\sim \mathbb{R}^{C}$. Additionally, during training, there is a load imbalance issue~\cite{fedus2022switch} among the expert layers: most inputs are dispatched to a small number of experts, while many other experts can not get sufficiently trained. To address this issue, we follow previous work~\cite{fedus2022switch} and add a balance loss to the training objective, encouraging a balanced load across the expert layers. The final loss function of DEMONet can be expressed as the combination of the recognition task loss and a load balancing loss:

\begin{equation}
\begin{aligned}
    \mathcal{L}
    &=\mathcal{L}_\mathrm{task} + \alpha \cdot \mathcal{L}_\mathrm{balance} \\
    &=CE(z_i, y_i) + \alpha N \cdot \sum_{i=1}^{N}{f_i} \cdot \sum_{i=1}^{N}{p^{T}_i}, \\
\end{aligned}
\end{equation}

where the $CE$ represents the cross-entropy loss, and the $f_i\sim \mathbb{R}^{N}$ denotes the number of inputs assigned to each expert. $\sum_{i=1}^{N}{f_i}$ signifies the total number of inputs in a batch assigned to each expert, while $\sum_{i=1}^{N}{p_i}$ represents the sum of probabilities of inputs in a batch assigned to each expert. The balance loss $\mathcal{L}_\mathrm{balance}$ in Equation (7) encourages a uniform routing assignment, as it is minimized under a uniform distribution for $f_i$ and $p_i$. To control the impact of the balance loss during training, a hyper-parameter $\alpha$ is introduced as a multiplicative coefficient for the loss. Throughout this study, we set the value of $\alpha$ to $10^{-2}$ by default, which is large enough to ensure load balancing while remaining small enough not to overwhelm the primary cross-entropy objective. The final loss $\mathcal{L}$ can be minimized when DEMONet makes accurate predictions with a uniform distribution for expert assignments. During the training process, the parameters of the routing layer, activated expert layers, and the backbone network are updated based on the gradient that minimizes $\mathcal{L}$, while the parameters of the cross-temporal VAE and inactive expert layers do not participate in the updates.

\begin{table*}[htbp]
\normalsize
    \centering
    \caption{The detailed network structure of DEMONet with cross-temporal VAE. The term ``Conv2d'' refers to the 2D convolution layer, while ``ConvTranspose2d'' represents the 2D transposed convolution layer. ``expert num'' corresponds to the number of expert layers, and ``out dim'' indicates the number of classes to be predicted.}
	\scalebox{0.75}{\begin{tabular}{ll}
		\hline
		  Module&Network layer\\ 
            \hline
            Encoder of VAE&Conv2d(in channel=1, out channel=64, kernel size=4, stride=2, padding=1)\\ 
             &Batch Normalization 2d(num features=64)\\ 
             &ReLU()\\ 
             &Conv2d(in channel=64, out channel=64, kernel size=4, stride=2, padding=1)\\ 
             &Batch Normalization 2d(num features=64)\\ 
             &ReLU()\\
             &Conv2d(in channel=64, out channel=64, kernel size=5, stride=1, padding=0)\\ 
             &Batch Normalization 2d(num features=64)\\ 
             &ReLU()\\
             &Conv2d(in channel=64, out channel=256, kernel size=3, stride=1, padding=0)\\ 
             &Batch Normalization 2d(num features=256)\\ 
             \hline
            Decoder of VAE&ConvTranspose2d(in channel=128, out channel=64, kernel size=3, stride=1, padding=0)\\ 
             &Batch Normalization 2d(num features=64)\\ 
             &ReLU()\\
             &ConvTranspose2d(in channel=64, out channel=64, kernel size=5, stride=1, padding=0)\\ 
             &Batch Normalization 2d(num features=64)\\ 
             &ReLU()\\
             &ConvTranspose2d(in channel=64, out channel=64, kernel size=4, stride=2, padding=1)\\ 
             &Batch Normalization 2d(num features=64)\\ 
             &ReLU()\\
             &ConvTranspose2d(in channel=64, out channel=1, kernel size=4, stride=2, padding=1)\\
             &Tanh()\\
             \hline
            Routing layer&Linear(in features=1172, out features=expert num)\\
            \hline
            Expert layer&Conv2d(in channel=1, out channel=64, kernel size=7, stride=2, padding=3)\\ 
             &Batch Normalization 2d(num features=64)\\ 
             &ReLU()\\ 
             &Max pooling(kernel size=3, stride=2, padding=1)\\
            \hline
            Backbone network&Basic block(64,64), Basic block(64,64)\\
             &Basic block(64,128),Basic block(128,128)\\
             &Basic block(128,256), Basic block(256,256)\\
             &Basic block(256,512), Basic block(512,512)\\
             &Adaptive average pooling(output size=(1, 1))\\
             &Linear(in features=512, out features=kind num)\\
            \hline
            Basic block(in dim, out dim) & Conv2d(in dim, out dim, kernel size=3, padding=1)\\
            &Batch Normalization 2d(num features=out dim)\\
            &ReLU()\\
            &Conv2d(out dim, out dim, kernel size=3, padding=1)\\
            &Batch Normalization 2d(num features=out dim)\\
		\hline
        \label{tab_arch}
	\end{tabular}}
\end{table*}

\subsection{Detailed Architectures}

This subsection provides a detailed description of the architectures used in our proposed DEMONet with cross-temporal VAE, as shown in Table~\ref{tab_arch}. In addition,  the detailed model architectures are visualized in Figure~\ref{fig2} (b).

The cross-temporal VAE consists of an encoder and a decoder. The encoder consists of several sequential layers, including four convolutional layers, four batch normalization (BN) layers, and three ReLU layers. Conversely, the decoder can be seen as the inverse operation of the encoder and also consists of several sequential layers. It includes four transposed convolution layers\footnote{We implemented the transposed convolution layer through torch.nn.ConvTranspose2d() in PyTorch.}, four BN layers, and three ReLU layers. The transposed convolution is an upsampling operation that first pads zero values between pixels of the input feature maps and then performs a standard convolution operation on the padded feature maps. At the end of the decoder, there is a Tanh layer, which is used to normalize the output value of the decoder to the range $[-1, 1]$.

In addition to VAE, other components of DEMONet include a routing layer, several expert layers, and a backbone network. The routing layer is a simple linear layer with the input dimension matching the dimension of the 1D DEMON spectra, and the output dimension corresponding to the number of expert layers. The decision to use such a straightforward structure for the routing layer is driven by the fact that it only needs to focus on low-level features of the 1D DEMON spectra, such as spectral peaks and harmonic periods. The simple structure not only brings lower computational costs but also reduces the risk of overfitting. The multiple expert layers share an identical structure, which includes a $7\times7$ convolutional layer, a BN layer, a ReLU layer, and a $3\times3$ max pooling layer. The backbone network follows the structure of ResNet-18~\cite{he2016deep} and  mainly consists of four residual layers. Each residual layer contains a stack of two basic blocks, which consist of two 3×3 convolutional layers, two BN layers, a ReLU layer, and a skip connection (as illustrated in Figure~\ref{fig2}). Additionally, the backbone network also includes an adaptive average pooling layer and a linear layer. The output dimension of the linear layer is determined by the number of categories to be predicted for the given task.

\section{Experiment Setup}

\subsection{Datasets}
In this study, we used three underwater ship-radiated noise datasets of different scales. The detailed information is listed in Table~\ref{tab1} and the following paragraphs:

1. DeepShip\footnote{The dataset is available at https://github.com/irfankamboh/DeepShip.}~\cite{irfan2021deepship} is a large-scale public underwater acoustic benchmark dataset, which contains 613 real-world underwater recordings with a sampling rate of 32000 Hz. The total duration is 47.07 hours, and the length of recordings ranges from 6 seconds to 25 minutes. The dataset includes 265 different types of ships from four classes: cargo ships, passenger ships, oil tankers, and tug boats.

2. Our private dataset - DTIL~\cite{ren2019feature} was collected from Thousand Island Lake, Hangzhou, China in 2019. The dataset is composed of 39 recordings with a uniform duration of 15 minutes and a sampling rate of 17067 Hz. It includes 300 minutes of speedboat recordings and 285 minutes of experimental vessel recordings with multiple sources of interference.

3. ShipsEar\footnote{The dataset is available at http://atlanttic.uvigo.es/underwaternoise.}~\cite{santos2016shipsear} is an open-source database of underwater recordings of ship-radiated sounds. The recordings were collected in different areas of the Atlantic coast of Spain between 2012 and 2013. The dataset is composed of 90 recordings with a sampling rate of 52734 Hz and lengths ranging from 15 seconds to 10 minutes. The total duration of recordings is approximately three hours. It includes 11 different types of ship sounds and 1 type of natural noise. In this study, we select a subset of 9 types of sounds (dredgers, fish boats, motorboats, mussel boats, ocean liners, passenger ships, ro-ro ships, sailboats, and natural noise) for the recognition task since the other three categories (pilot ships, trawlers, tug boats) lack sufficient samples to support the train-validation-test split.

It is worth mentioning that our experiments primarily revolved around DeepShip and DTIL datasets, as they provide an ample amount of training data. The ShipsEar dataset was only employed to assess the performance of DEMONet in data scarcity scenarios (see Section 5.6).

\begin{table*}[htbp]
\normalsize
    \centering
    \caption{Information of the three datasets. ``sr'' represents the sampling rate and ``d()'' represents the dimension of features.}
	\scalebox{1}{\begin{tabular}{lccclll}
		\hline
		  dataset  & duration (hours) & sr (Hz) & pass bands (Hz)&
            d(STFT)&d(Mel)&d(CQT)\\
            \hline
            Deepship & 47.07 & 32000 & 10-8000&
            1199,400& 1199,300&899,289\\
            DTIL  & 10.25 & 17067 & 10-2000&
            1199,95&1199,300&899,229\\
            ShipsEar  & 2.94 & 52734 & 100-26360&
            1199,1313&1199,300&899,340\\   
		\hline
        \label{tab1}
	\end{tabular}}
\end{table*}

\subsection{Data division}
For this study, we divided each signal record into 30-second segments with a 15-second overlap. To mitigate information leakage, we ensure that segments in the training and test sets do not belong to the same audio track. This restriction allows us to evaluate our model's generalization performance.

Most previous studies on underwater acoustic recognition tasks did not release their train-test split, making it difficult to make a fair comparison. To address this limitation, we released a carefully selected train-test split~\cite{xie2024unraveling} for DeepShip and ShipsEar (see Appendix C) in this paper, with 15\% of the data in the training set randomly chosen as the validation set. We hope that our released split can serve as a reliable benchmark for future work aiming to make fair comparisons.

\subsection{Parameters Setup}
During framing, we set the frame length to 50ms and the frame shift to 25ms according to the preliminary experiments outlined in Appendix B. The number of Mel filter banks was set to 300 by default. Regarding the CQT kernel, we uniformly set the octave resolution and time resolution to 30 as the default values. The maximum (minimum) frequency, denoted as $f_{max}$ ($f_{min}$) for the CQT transform was determined by the passband boundaries specific to each dataset (refer to Table~\ref{tab1}).

During training, we used the AdamW~\cite{loshchilov2018decoupled} 
optimizer. The maximum learning rate was set to 5$\times 10^{-3}$ and the weight decay was set to $10^{-3}$ for all experiments. We use dthe cosine annealing schedule for learning rate decay, with a warm-up epoch of 5. All models were trained for 200 epochs on Nvidia A10 GPUs. The CUDA version is 11.4, the Python version is 3.8.8 and the Pytorch version is 1.9.0.

\section{Results and Analysis}
In this study, the uniform evaluation metric adopted for the multi-class recognition task was the accuracy rate, which can be calculated by dividing the number of correctly predicted samples by the total number of samples. Given the limited quantity of signal recordings in the test set, there exist cases where multiple groups of experiments produced the same file-level accuracy.  To address this, we presented the results at the segment level (30 seconds) rather than the file level. Furthermore, to mitigate the influence of randomness, this study utilized two distinct random seeds (123, 3407) for all implemented experiments. The reported results provided the average accuracy and variations across random seeds. 

\begin{figure*}[ht]
    \centering
    \includegraphics[width=0.85\linewidth]{figs-demonet/fig3.pdf}
    \caption{Preliminary Experiments on selecting backbone model architectures and input features.}
    \label{fig_pre}
    \vspace{-2px}
\end{figure*}

\subsection{Preliminary Experiments}
We conducted preliminary experiments on DeepShip and DTIL to search for the optimal parameter setups (refer to Appendix A and B), backbone model architectures, and input features. Besides, the preliminary experiment also involved verifying the effects of directly integrating DEMON features into recognition models.

To select the optimal backbone model architectures, we implemented several models for comparison, including SVM~\cite{yang2016underwater}, fully conv networks (FCN)~\cite{liu2021underwater}, MobileNet v3~\cite{howard2019searching}, ResNet~\cite{he2016deep}, SE-ResNet~\cite{hu2018squeeze}, and ResNet with attention~\cite{xie2022underwater}. To ensure a fair comparison, all experiments uniformly used the STFT amplitude spectrogram as the input feature. As shown in Figure~\ref{fig_pre} (a), ResNet-18 exhibited superior performance compared to other models on both datasets. Particularly noteworthy is the fact that ResNet-18 surpassed other ResNet-based variants, including SE-ResNet and ResNet with attention. This observation suggests that the structure and complexity of ResNet-18 are well-suited for the task and can serve as a reliable backbone network. Furthermore, we compared the performance of models using different input features. For a fair comparison, ResNet-18 was used as the backbone architecture for all experiments. The results in Figure~\ref{fig_pre} (b) indicate that the 2-D DEMON spectra performed the worst, with a recognition accuracy approximately 20\% lower than that of spectrograms on DeepShip. This suggests that the DEMON spectra could not provide sufficient class-specific discriminative patterns for the recognition task. Among the other three time-frequency features, the CQT spectrogram performed the best. According to our previous research~\cite{xie2023guiding}, spectrogram-based recognition models typically focus on the low-frequency line spectrum (frequency domain) and high-frequency modulation information (temporal domain) of signals. The CQT spectrogram, with its high frequency resolution at low frequencies and high temporal resolution at high frequencies, is well-suited as a feature for spectrogram-based recognition systems. Consequently, we adopted the CQT spectrogram as the default feature for subsequent experiments.

\begin{table*}[htbp]
\normalsize
    \centering
    \caption{Preliminary experiments on feature fusion or model ensemble. All methods used CQT spectrograms as one of the features by default. E.g., ``+DEMON spec'' represents implementing feature fusion based on CQT spectrograms and DEMON spectra.}
	\scalebox{0.9}{\begin{tabular}{llccc}
		\hline
		  Features &  Methods & DeepShip (\%)& DTIL (\%)& Benefits (\%)\\
            \hline
            CQT spec & - & 77.15$\pm$1.16& 97.18$\pm$0.14& -,-\\
            +Amplitude spec & ICL(feature fusion)~\cite{xie2023guiding} &77.72$\pm$1.00&97.51$\pm$0.10&+0.57,+0.33\\
            +DEMON spec & ICL(feature fusion) &74.45$\pm$0.47&96.78$\pm$0.26&- 2.70, - 0.40\\
            +Mel spec+Amplitude spec & Majority voting(model ensemble) &77.50&97.22&+0.35,+0.04\\
            +Mel spec+DEMON spec & Majority voting(model ensemble) &77.44&97.18&+0.29,+0.00\\
		\hline
        \label{tab_demon}
	\end{tabular}}
\end{table*}

In addition, we conducted preliminary experiments using feature fusion or model ensemble methods~\cite{xie2023guiding,yang2016underwater} to examine the effects of directly integrating DEMON features into recognition models. The results in Table~\ref{tab_demon} demonstrate that it is challenging to derive benefits from directly integrating DEMON features. Whether based on feature fusion or model integration methods, the benefits provided by DEMON features were lower than those offered by the STFT amplitude spectrogram. Particularly for the feature fusion method, the performance of the DEMON-based model (74.45\% on DeepShip) was even significantly worse than the baseline (77.15\% on DeepShip). Our analysis indicates that this detrimental effect is due to the lack of class-specific discriminative patterns in DEMON features, which may introduce inappropriate inductive bias and negatively influence the model's performance. The experimental results convincingly demonstrated the necessity of our study, specifically why we need to design the DEMONet to utilize the DEMON features in an indirect manner.

\subsection{Comparative Experiments}

\begin{table*}[htbp]
\normalsize
    \centering
    \caption{Comparative experiments of DEMONet and advanced underwater acoustic target recognition methods. ``Para Num'' indicates the number of parameters of the model. ``Baseline'' here represents the method using CQT spectrogram and ResNet-18.}
	\scalebox{0.9}{\begin{tabular}{lccc}
		\hline
		  Methods &  DeepShip(\%) & DTIL(\%) & Para Num (MB)\\
            \hline
            TDNN and WPCS~\cite{ren2019feature} &73.20$\pm$0.49&95.20$\pm$0.11&2.078\\
             AGNet~\cite{xie2022adaptive} &76.99$\pm$0.10&95.71$\pm$0.05&28.634\\
             SIR-ResNet~\cite{xie2023advancing} &77.27$\pm$0.98&97.04$\pm$0.14&10.657\\
             ICL(CQT\&Amplitude spec)~\cite{xie2023guiding}
             &77.72$\pm$1.00&97.51$\pm$0.10&21.811\\
             Convolutional-MoE~\cite{xie2024unraveling}
             &78.22$\pm$1.40&97.46$\pm$0.42&10.672\\
             \hline
             Baseline &77.15$\pm$1.16& 97.18$\pm$0.14&10.657\\
            DEMONet - 2 expert&77.78$\pm$0.20 &97.18$\pm$0.00&10.662\\
            DEMONet - 5 expert &80.03$\pm$0.91&97.46$\pm$0.43&10.671\\
            DEMONet - 8 expert&79.13$\pm$0.59 & 97.20$\pm$0.15&10.687\\
            Cross-Temporal VAE+DEMONet - 2 expert&78.14$\pm$0.04 &97.49$\pm$0.11&0.535+10.662\\
            Cross-Temporal VAE+DEMONet - 5 expert &\textbf{80.45$\pm$0.67}&\textbf{97.88$\pm$0.25}&0.535+10.671\\
            Cross-Temporal VAE+DEMONet - 8 expert&80.03$\pm$0.15 & 97.67$\pm$0.00&0.535+10.687\\
            \hline
        \label{tab4}
	\end{tabular}}
\end{table*}

In this subsection, we presented the results of comparative experiments conducted against advanced underwater acoustic recognition methods. In this study, we implemented several existing advanced underwater acoustic recognition methods (TDNN and WPCS~\cite{ren2019feature}, AGNet~\cite{xie2022adaptive}, SIR-ResNet~\cite{xie2023advancing}, ICL~\cite{xie2023guiding}, Convolutional-MoE~\cite{xie2024unraveling}) on DeepShip and DTIL for comparisons. As depicted in Table~\ref{tab4}, the results show that our DEMONet, especially with five expert layers, can achieve steady improvement on both datasets compared to the baseline. Moreover, DEMONet introduced only up to 0.03 MB of additional parameters compared to the baseline method, which brought almost no additional computing consumption. Furthermore, the incorporation of cross-temporal VAE can further enhance the performance of DEMONet to achieve a recognition accuracy of 80.45$\pm$0.67\% on DeepShip at the cost of 0.535 MB of additional parameters, reaching the current state-of-the-art performance. On DTIL, DEMONet with cross-temporal VAE can also achieve a recognition accuracy of 97.88$\pm$0.25\%, surpassing all other advanced methods.

Moreover, we investigated the effect of the number of experts on DEMONet. As depicted in Table~\ref{tab4}, DEMONet with five experts demonstrated the best performance on both DeepShip and DTIL. Setting the number of expert layers to five appeared to be an appropriate practice since more expert layers may lead to additional computational consumption and a higher risk of overfitting, while fewer expert layers may result in reduced model capacity and underutilization of the multi-expert system's potential. Therefore, in subsequent experiments, DEMONet was configured with five expert layers by default.

It is worth noting that DEMONet exhibited a significant improvement over the baseline on DeepShip, whereas the improvement on DTIL was relatively limited. Since the signals in Deepship are abundant in quantity and exhibit clear modulation rhythms, DEMONet can get fully trained with appropriate routing assignments and successfully utilize the physical characteristics to perceive underwater signals. In contrast, DTIL has a simpler acquisition environment and includes only two categories of targets. Therefore, the model does not heavily rely on physical characteristics to achieve satisfactory performance on DTIL. This finding suggests that the benefits brought by DEMONet are data-dependent. Adequate data quantity, complex data distribution, and clear modulation spectra can stimulate DEMONet's capabilities.

\subsection{Ablation Experiments}

\begin{table*}[htbp]
\normalsize
    \centering
    \caption{The results of the ablation experiments. All models were configured with five expert layers by default. ``VAE $\checkmark$+CTVAE $\times$'' represents applying vanilla VAE to reconstruct DEMON spectra, while ``VAE $\checkmark$+CTVAE $\checkmark$'' represents using cross-temporal VAE to reconstruct DEMON spectra.}
	\scalebox{0.9}{\begin{tabular}{ccccc}
		\hline
	\quad VAE\quad &\quad Cross-Temporal VAE\quad & \quad load balancing loss\quad &\quad DeepShip(\%)\quad &\quad DTIL(\%)\quad\\
            \hline
        $\times$ & $\times$ &$\times$ & 
         79.54$\pm$1.87&97.32$\pm$0.43 \\
        $\times$ & $\times$ &$\checkmark$ & 
         80.03$\pm$0.91&97.46$\pm$0.43 \\
        $\checkmark$ & $\times$ &$\times$ & 
         79.49$\pm$1.33 &  97.37$\pm$0.00 \\
         $\checkmark$ & $\checkmark$ &$\times$ & 
         79.79$\pm$1.05 &97.88$\pm$0.25\\
        $\checkmark$ & $\checkmark$ &$\checkmark$ & 
         80.45$\pm$0.67&97.88$\pm$0.25 \\
        
        \hline
        \label{tab_abl}
        \vspace{-2mm}
	\end{tabular}}
\end{table*}

The results of ablation experiments on VAE and load balancing loss were presented in Table~\ref{tab_abl}. Firstly, the experimental results demonstrated the beneficial effect of the cross-temporal VAE, which has been introduced in Section 5.2. Secondly, we observed that the load balancing loss can consistently improve model performance. It helps address the issue where certain expert layers are not adequately trained due to unbalanced input assignments. Additionally, ablation experiments were conducted on the cross-temporal mechanism. The results indicated that the vanilla VAE without the cross-temporal mechanism can not yield improvements and may even have negative effects (e.g., drop from 79.54\% to 79.49\% on DeepShip). This phenomenon arises from the limitation of the vanilla VAE, which is unable to effectively diminish noise or spurious modulation spectra in DEMON features, and has the potential risk of losing characteristics spectral lines. The result strongly supports the necessity of our proposed cross-temporal mechanism. In the next subsection, we will continue to delve into the effect of cross-temporal VAE.

\subsection{Analyses on Cross-Temporal VAE}

\begin{figure*}[htbp]
    \centering
    \includegraphics[width=\linewidth]{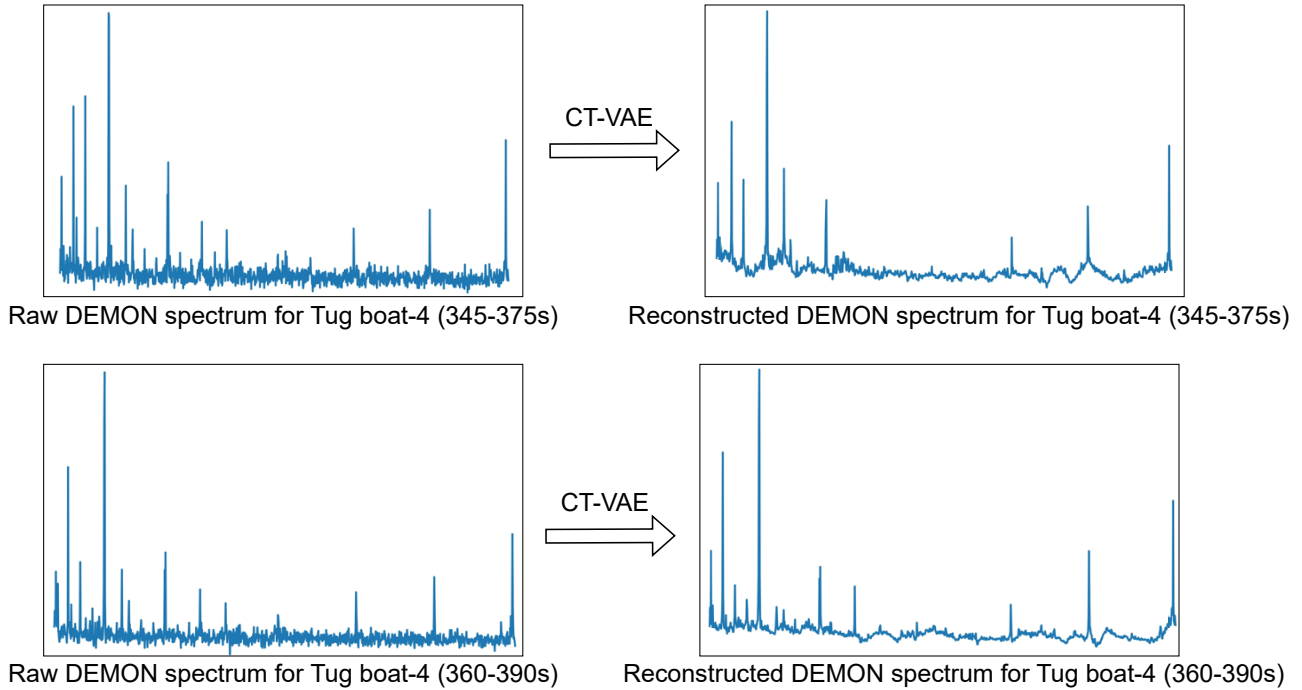}
    \caption{A comparison between the raw 1-D DMEON spectra (left) and the reconstructed 1-D DEMON spectra using cross-temporal VAE (right).}
    \label{fig_dmeon}
    \vspace{-2px}
\end{figure*}

In this subsection, we delved deeper into the effects of cross-temporal VAE on DEMONet and conducted a thorough analysis. Firstly, we showcased the raw 1-D DMEON spectra alongside the reconstructed 1-D DEMON spectra in Figure~\ref{fig_dmeon} for comparison. The figure reveals that the continuous spectrum in the reconstructed 1-D DEMON spectra is effectively suppressed, thereby reducing interference caused by environmental noise or non-target sound sources. we observed a slight variation in the number of spectral lines between the raw DEMON spectra of different segments for the same target, potentially due to interferences in the marine environment. Whereas, after the reconstruction of cross-temporal VAE, the spectral lines in the 345-375s segment were reduced and displayed a closer resemblance to those in the 360-390s segment. This observation preliminarily indicates that cross-temporal VAE can reduce noise and interference in DEMON spectra, enabling a focus on target-related physical characteristics.

Furthermore, to quantify the similarity of DEMON spectra from the same signal recording across different periods, we employed cosine similarity as the metric. We randomly selected several segments from the same signal within each DeepShip category and compared the cosine similarity of their 1-D DEMON spectra. The results, shown in Table~\ref{tab_cossim}, provide evidence that DEMON spectra reconstructed by cross-temporal VAE have higher consistency across different periods. This consistency aligns with the time-invariant nature of physical characteristics, affirming that cross-temporal VAE can reconstruct DEMON spectra with reduced interference while preserving robust patterns. Such reconstruction is beneficial for the routing layer to better comprehend input DEMON spectra and make appropriate routing assignments.

\begin{table*}[htbp]
\normalsize
    \centering
    \caption{The cosine similarity of 1-D DEMON spectra belonging to the same signal recording across different periods. ``Raw Similarity'' denotes the cosine similarity between the raw 1-D DEMON spectra from different periods, while `` Recon Similarity'' represents the cosine similarity between the reconstructed 1-D DEMON spectra from different periods.}
	\scalebox{0.9}{\begin{tabular}{llcc}
		\hline
	Segments $wav_A$ & Segments $wav_B$ &Raw Similarity&Recon Similarity\\
         \hline
         Cargo ship-101 (165-195s)&Cargo ship-101 (180-210s) & 0.9867& 0.9885 \\
         Cargo ship-101 (165-195s)&Cargo ship-101 (195-225s) & 0.9826& 0.9866 \\
         Cargo ship-101 (180-210s)&Cargo ship-101 (195-225s) & 0.9868& 0.9882 \\
         Average&&0.9854&0.9878\\
        \hline
        Passenger ship-47 (75-105s)&Passenger ship-47 (90-120s) & 0.9454& 0.9687 \\
         Passenger ship-47 (75-105s)&Passenger ship-47 (105-135s) & 0.9403& 0.9652 \\
         Passenger ship-47 (90-120s)&Passenger ship-47 (105-135s) & 0.9660& 0.9808 \\
         Average&&0.9506&0.9716\\
        \hline
        Tanker-216 (165-195s)&Tanker-216 (180-210s) & 0.9783& 0.9838 \\
         Tanker-216 (165-195s)&Tanker-216 (195-225s) & 0.9721& 0.9807 \\
         Tanker-216 (180-210s)&Tanker-216 (195-225s) & 0.9813& 0.9864 \\
         Average&&0.9772&0.9836\\
        \hline
        Tug boat-4 (345-375s)&Tug boat-4 (360-390s)
        &0.9874&0.9883\\
        Tug boat-4 (345-375s)&Tug boat-4 (375-405s)
        &0.9812&0.9862\\
        Tug boat-4 (360-390s)&Tug boat-4 (375-405s)
        &0.9818&0.9867\\
        Average&&0.9835&0.9871\\
        \hline
        
        \label{tab_cossim}
        \vspace{-2mm}
	\end{tabular}}
\end{table*}

\subsection{Routing Assignment}

\begin{figure*}[htbp]
    \centering
    \includegraphics[width=\linewidth]{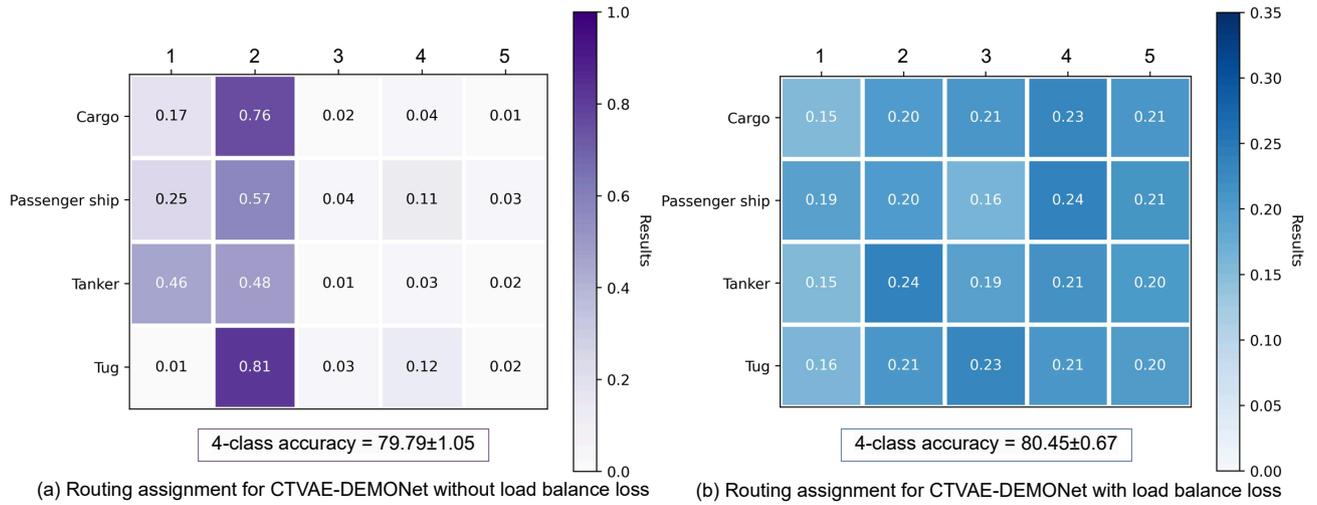}
    \caption{The routing assignment for DEMONet on the DeepShip training set. The horizontal axis scale represents the ID of the expert layer, while the vertical axis scale represents the target type. Each grid cell displays the proportion of the number of targets assigned to a certain expert layer to the total number of targets of that specific category.}
    \label{fig_route}
    \vspace{-2px}
\end{figure*}

For our proposed multi-expert structure, the routing assignment of expert layers is also worthy of attention. We have visualized the routing assignment on the DeepShip training set and presented it in Figure~\ref{fig_route}. Figure~\ref{fig_route} (a) clearly demonstrates the presence of load imbalance when the load balancing loss was not applied. Notably, the majority of inputs were assigned to $\mathrm{Expert_2}$, while $\mathrm{Expert_3}$ and $\mathrm{Expert_5}$ did not receive sufficient training samples. Despite this, the model still managed to achieve a recognition accuracy of 79.79$\pm$1.05\% on the DeepShip test set, primarily due to the fact that most samples in the test set were also assigned to the well-trained $\mathrm{Expert_2}$. However, in practical scenarios, there may be numerous unseen signals that are unsuitable for processing with the specialized knowledge of $\mathrm{Expert_2}$. Consequently, the model's performance may significantly degrade as most experts are highly underfitting, resulting in poor generalization capabilities. By incorporating the load balancing loss (Figure~\ref{fig_route} (b)), the recognition accuracy on DeepShip demonstrates a slight improvement. Moreover, the routing assignment strikes a good balance, allowing each expert to receive comprehensive training and effectively mitigating the aforementioned risks.

\subsection{Performance on ShipsEar}
To assess the performance of DEMONet in a data scarcity scenario, we conducted validation experiments and compared our method with other advanced techniques on the ShipsEar dataset. The Shipsear dataset comprises a total data duration of less than 3 hours, with fewer than 500 training samples at the 30-second level in the training set. The results, as presented in Table~\ref{tab_ShipsEar}, demonstrate that our proposed cross-temporal VAE and DEMONet can yield significantly better recognition results compared to the ResNet-18 baseline. However, there remains a certain gap when compared to the current state-of-the-art method~\cite{xie2024unraveling}. The limited amount of data in the ShipsEar dataset poses a challenge for training the expert layers, with each expert layer often receiving less than 100 training samples in a single epoch. Consequently, DEMONet's capabilities may not be fully utilized. Additionally, we have observed that the signals in ShipsEar lack clear modulation information, resulting in DEMON spectra containing numerous interference spectral lines. Even with the assistance of the cross-temporal VAE, DEMONet struggles to accurately focus on the physical characteristics of targets. This difficulty hampers the model's ability to establish a reasonable routing assignment, thereby limiting its recognition performance. In essence, DEMONet is a data-driven approach that relies on sufficient data to fully leverage its ability to process data at a fine-grained level.

\begin{table*}[htbp]
\normalsize
    \centering
    \caption{The performance of DEMONet and several advanced methods on ShipsEar.}
	\scalebox{0.9}{\begin{tabular}{llcc}
		\hline
		  Feature& Model&9-class Accuracy (\%)\\
            \hline
            STFT amplitude spectrogram&ResNet-18~\cite{he2016deep}& 75.24\\
             &SIR-ResNet (+LMR)~\cite{xu2023underwater} &81.90 (82.97)\\
             &Convolutional-MoE~\cite{xie2024unraveling}&86.21\\
             &Cross-Temporal VAE+DEMONet&84.92$\pm$2.16\\
             \hline
            Mel spectrogram&ResNet-18& 77.14\\
             &SIR-ResNet (+LMR)&82.76 (83.45)\\
             &Convolutional-MoE&85.35\\
             &Cross-Temporal VAE+DEMONet&84.05$\pm$0.43\\
             \hline
             CQT spectrogram&ResNet-18& 73.33\\
             &SIR-ResNet (+LMR)&75.86 (82.76)\\
             &Convolutional-MoE&82.76\\
             &Cross-Temporal VAE+DEMONet&81.04$\pm$1.72\\
             \hline
        \label{tab_ShipsEar}
	\end{tabular}}
\end{table*}

\section{Conclusions}
In this study, we propose a multi-expert network -- DEMONet, to integrate physical characteristics into recognition systems to compensate for the lack of robustness. To mitigate the noise and spurious modulation spectra in DEMON spectra, we propose a cross-temporal variational autoencoder. Furthermore, we employ a load balancing loss to tackle the issue of load imbalance. Detailed experiments were conducted to validate the effectiveness of our proposed methods. It is worth mentioning that our DEMONet can achieve state-of-the-art on the DeepShip and DTIL datasets.

Despite the promising performance on the two datasets, our method still has some limitations and room for improvement. Firstly, DEMONet's performance may not be optimal when the available data is limited or when the modulation information is indistinct. Additionally, this study solely utilizes DEMON spectra to provide physical characteristics. In subsequent work, we aim to focus on exploring and developing recognition systems that leverage multiple physical characteristics or environmental factors to provide more comprehensive insights.

\section*{Acknowledgements}
This research is supported by the IOA Frontier Exploration Project (No. ZYTS202001) and Youth Innovation Promotion Association CAS.

\section*{Declaration of Generative AI and AI-Assisted Technologies in the Writing Process}
During the preparation of this study, we used $ChatGPT$ in order to check the English grammar and word spelling. After using this service, we reviewed and edited the content as needed and take full responsibility for the content of the publication.

\appendix
\section{Selection of Passbands}
\label{sec:appendix1}

\begin{table*}[htbp]
\normalsize
    \caption{The preliminary experiment on different passbands. The frame length/shift is set to 50ms/25ms by default.}
    \centering
	\scalebox{0.9}{
	\begin{tabular}{lccclc}
        \hline
	Pass bands(Hz) & DeepShip(\%) & DTIL(\%)&\quad \quad \quad \quad &Pass bands(Hz)& ShipsEar(\%)\\
	\hline
	10-2k & 72.54 & \textbf{96.26} &&100-8k & 60.69 \\
        10-4k& 74.37& 95.93&&100-12k & 64.14  \\
        10-8k& \textbf{74.68} & 95.08&&100-16k & 70.90 \\
	10-12k& 74.53 & -&&100-24k & 74.34\\
        10-16k& 73.66 & -&&100-26367 & \textbf{75.24} \\
        \hline
        \label{tabA}
	\end{tabular}}
\end{table*}

To eliminate low-frequency interference and redundant high-frequency components, band-pass filtering was applied during the preprocessing stage. We implemented preliminary experiments to determine the optimal passband ranges for three datasets. The candidate frequency ranges for the passbands were set based on the Nyquist theorem and the cutoff frequencies of the hydrophones~\cite{ren2019feature} and preamplifiers~\cite{santos2016shipsear}. For all experiments, we uniformly utilized the STFT amplitude spectrogram as the input feature and employed the ResNet-18 model. The random seed was set to 123 for consistency across all experiments. The recognition results with different passbands on the three datasets are presented in Table~\ref{tabA}. Based on these results, we selected a passband of 10-8000Hz for DeepShip, 10-2000Hz for DTIL, and 100-26367Hz for ShipsEar.

\section{Selection of Frame Length and Frame Shift}
\label{sec:appendix2}

The selection of frame length and frame shift can impact the temporal and frequency resolution of acoustic features, thus influencing the performance of the recognition system. We conducted experiments to evaluate the recognition results using different frame lengths and frame shifts, and the corresponding results are presented in Table~\ref{tabB}. For all experiments, we consistently employed the STFT amplitude spectrogram as the input feature and utilized the ResNet-18 model. The random seed was set to 123. Based on the results, we observed that setting the frame length to 50ms and the frame shift to 25ms consistently produced the best recognition performance across all three databases.

\begin{table*}[htbp]
\normalsize
    \caption{The preliminary experiment on frame length and frame shift.}
    \centering
	\scalebox{0.9}{
	\begin{tabular}{llccc}
        \hline
	Frame length/Frame shift  &DeepShip & DTIL&ShipsEar\\
	\hline
	30ms/15ms& 73.48 & 95.10&  72.78 \\
        50ms/25ms& \textbf{74.68} & \textbf{96.26}&  \textbf{75.24} \\
        100ms/50ms& 74.25& 95.77 & 72.78 \\
	150ms/75ms& 71.97 & 95.31& 74.56   \\

        \hline
        \label{tabB}
	\end{tabular}}
\end{table*}

\section{Train-Test Split for DeepShip and ShipsEar}
\label{sec:appendix3}

We have noticed that the majority of prior studies on underwater acoustic recognition tasks did not disclose their division of training/test sets, which poses a challenge when attempting to make a fair comparison. In this study, we provided our carefully selected train-test split for DeepShip and ShipsEar (refer to Table~\ref{tabC1} and~\ref{tabC2}). We anticipate that our train-test split can serve as a reliable benchmark for future studies that aim to realize fair comparisons.\footnote{The train-test-split for ShipsEar is also released on\\ https://github.com/xy980523/ShipsEar-An-Unofficial-Train-Test-Split.}.

\begin{table*}[ht]
\normalsize
    \caption{\label{tabC2} Train-test split for DeepShip. The ``ID'' in the table refers to the ID of the .wav file in the dataset.}
    \centering
	\scalebox{0.85}{
	\begin{tabular}{lll}
        \hline
	Category& ID in Training set & ID in Test set\\
	\hline
    Cargo ship& Else & 01,02,04,05,18,30,32,35,40,48,56,62,\\
    & &63,67,68,72,74,79,83,91,92,93,95,97,\\
    & &100,104\\

    \hline
    Passenger ship& Else&02,04,05,07,11,15,17,19,20,23,30,31,\\
    &&37,45,46,52,53,60,61,62,64,67,68,70,\\
    &&75,76,77,84,86,91,101,106,113,117,122,\\
    &&125,129,130,134,135,142,144,152,157,\\
    &&159,161,167,168,177,179,187,188,189\\

    \hline
    Oil tanker& Else&02,03,04,07,08,13,14,15,19,22,25,28,\\
    &&35,37,46,58,62,71,73,79,82,84,88,89,\\
    &&92,99,106,115,118,124,126,127,131,134,\\
    &&141,144,147,151,153,156,158,167,171,\\
    &&178,179,185,186,190,192,193,201,205,\\
    &&213,217,228,233\\
    \hline
    Tug boat& Else& 07,08,18,20,24,25,27,29,32,33,37,39,\\
    &&40,44,45,56,59,70\\
        \hline
	\end{tabular}}
\end{table*}

\begin{table*}[htbp]
\normalsize
    \caption{\label{tabC1} Train-test split for ShipsEar. The ``ID'' in the table refers to the ID of the .wav file in the dataset.}
    \centering
	\scalebox{0.9}{
	\begin{tabular}{lll}
        \hline
	Category& ID in Training set & ID in Test set\\
	\hline
	Dredger&80,93,94,96& 95 \\
    Fish boat&73,74,76&75 \\
    Motorboat&21,26,33,39,45,51,52,70,77,79&27,50,72 \\
    Mussel boat&46,47,49,66&48 \\
    Natural noise&81,82,84,85,86,88,90,91&83,87,92\\
    Ocean liner&16,22,23,25,69&24,71\\
    Passenger ship&06,07,08,10,11,12,14,17,32,34,36,38,40,&9,13,35,42,55,62,65\\
    &41,43,54,59,60,61,63,64,67& \\
    RORO ship&18,19,58&20,78\\
    Sailboat&37,56,68&57\\
        \hline
	\end{tabular}}
\end{table*}


\newpage

\printcredits

\bibliographystyle{cas-model2-names}
\bibliography{car-refs-demonet}





\end{document}